%% file: main.tex
\documentclass[sigplan,nonacm]{acmart}
\usepackage{booktabs}
\usepackage{xspace}
\usepackage{algorithm}
\usepackage[noend]{algpseudocode}

\usepackage{tabularx}
\usepackage[normalem]{ulem}
\usepackage{subcaption}
\usepackage[most]{tcolorbox}
\definecolor{tkbar}{RGB}{58,58,58}
\definecolor{tkfill}{RGB}{242,242,242}
\newtcolorbox{takeaway}[1]{%
  breakable=false, enhanced, arc=2.2pt, auto outer arc,
  colback=tkfill, colframe=tkbar, boxrule=0.6pt,
  title={\textbf{Takeaway #1}}, coltitle=white, colbacktitle=tkbar,
  fonttitle=\bfseries,
  lefttitle=7.9pt, righttitle=7.9pt, toptitle=3.7pt, bottomtitle=3.7pt,
  left=7.9pt, right=7.9pt, top=5.1pt, bottom=5.1pt, boxsep=0pt,
  before skip=7pt, after skip=7pt,
  fontupper=\bfseries}
\newcolumntype{L}[1]{>{\hsize=#1\hsize\raggedright\arraybackslash}X}

\newcommand{\runin}[1]{\noindent\textbf{#1.}\space}

\input{macros}

\input{results_macros}

\begin{document}
\brokenpenalty=100

\title{\lachesis: Lifetime-Aware KV Cache Placement for Agent Serving across HBM and High-Bandwidth Flash}

\author{Jaehoon Yang}
\affiliation{\institution{Seoul National University}\city{Seoul}\country{Republic of Korea}}
\email{jaehoon.yang@snu.ac.kr}

\author{Jeongmin Lee}
\affiliation{\institution{KAIST}\city{Daejeon}\country{Republic of Korea}}
\email{jeongminl@kaist.ac.kr}

\author{Haneul Park}
\affiliation{\institution{University of Illinois, Urbana-Champaign}\city{Urbana}\state{IL}\country{USA}}
\email{hnpark2@illinois.edu}

\author{Seung Yul Lee}
\affiliation{\institution{Seoul National University}\city{Seoul}\country{Republic of Korea}}
\email{triomphant1@snu.ac.kr}

\author{Nam Sung Kim}
\affiliation{\institution{University of Illinois, Urbana-Champaign}\city{Urbana}\state{IL}\country{USA}}
\email{nam.sung.kim@gmail.com}

\author{Jae W. Lee}
\affiliation{\institution{Seoul National University}\city{Seoul}\country{Republic of Korea}}
\email{jaewlee@snu.ac.kr}

\begin{abstract}
Large language model (LLM) serving is increasingly dominated by agentic workloads, in which agents and their sub-agents accumulate context as KV cache across many requests, consuming substantial memory. High-bandwidth flash (HBF) is a promising solution, providing an order of magnitude greater capacity at HBM-class read bandwidth, but its finite write endurance is the key limiting factor.
Our key insight is that KV cache should be placed across HBM and HBF by its lifetime. Placing shorter-lived data in HBM lets HBM absorb more of an agent run's writes and sends less of them to HBF.
As the lifetime of KV cache in agentic serving is dictated by the harness, the program that orchestrates the agents, we analyze its behavior and identify three axes along which lifetime diverges, temporal, structural, and \interworker.
Guided by these observations, we present \lachesis, a lifetime-aware KV cache placement layer between the agent harness and the serving engine. At write time, it places each segment in HBM or HBF according to its lifetime, and frees its blocks once the segment is no longer read. In trace-driven simulation, \lachesis extends HBF lifetime by \rEvalHbfLifetime over HBM-first placement, reaching \rEvalHbfAbsolute. Even under continuous 24$\times$7 operation at the full load a tight SLO admits, HBF outlasts its five-year warranty on the multi-agent trace.

\end{abstract}

\maketitle
\renewcommand{\shortauthors}{Yang et al.}

\input{contents/introduction}

\input{contents/background}

\input{contents/analysis}

\input{contents/placement}
\input{contents/evaluation}

\input{contents/related}

\input{contents/conclusion}

\phantomsection\label{paper:bodyend}

\bibliographystyle{ACM-Reference-Format}
\bibliography{references}

\appendix
\input{contents/appendix_corpus}
\input{contents/appendix_accounting}
\clearpage

\end{document}

%% file: macros.tex
\newcommand{\lachesis}{Lachesis\xspace}

\newcommand{\secref}[1]{\S\ref{#1}}

\definecolor{revorange}{HTML}{D95F02}
\definecolor{revblue}{HTML}{0072B2}

\definecolor{revgreen}{HTML}{009E73}

\definecolor{phgreen}{HTML}{009E73}
\newcommand{\placeholdervalues}{}

\definecolor{dummyblue}{RGB}{0,78,168}

\newcommand{\seg}{s}

\newcommand{\Vw}[1][\seg]{V_W(#1)}
\newcommand{\Vr}[1][\seg]{V_R(#1)}

\newcommand{\rhorw}[1][\seg]{\rho_{R/W}(#1)}
\newcommand{\rhorwbare}{\rho_{R/W}}

\newcommand{\tnorm}[1][\seg]{T_{\mathrm{norm}}(#1)}
\newcommand{\tnormbare}{T_{\mathrm{norm}}}

\newcommand{\Gdec}[1][\seg]{G_{\mathrm{dec}}(#1)}       %
\newcommand{\Nobs}[1][\seg]{N_{\mathrm{obs}}(#1)}       %
\newcommand{\lamtool}[1][\seg]{\Lambda_{\mathrm{tool}}(#1)} %
\newcommand{\lamuser}[1][\seg]{\Lambda_{\mathrm{user}}(#1)} %
\newcommand{\tpotref}{\mathrm{TPOT}_{\mathrm{ref}}}
\newcommand{\ttftref}{\mathrm{TTFT}_{\mathrm{ref}}}

\newcommand{\tnormeq}{%
  \begin{split}
    \tnorm = {}& \Gdec \, \tpotref + \Nobs \, \ttftref \\
               &{} + \lamtool + \lamuser
  \end{split}
}

\newcommand{\Lout}{L_{\mathrm{out}}}                    %
\newcommand{\Vkvtok}{V_{\mathrm{KV}}^{\mathrm{token}}}  %
\newcommand{\Nw}[1][\seg]{N_W(#1)}                      %
\newcommand{\Qset}[1][\seg]{\mathcal{Q}(#1)}            %

\newcommand{\shortlived}{short-lived context segment\xspace}
\newcommand{\longlived}{long-lived context segment\xspace}
\newcommand{\ladenseg}{reasoning-laden context segment\xspace}
\newcommand{\strippedseg}{reasoning-stripped context segment\xspace}
\newcommand{\leadseg}{lead-agent segment\xspace}
\newcommand{\subseg}{sub-agent segment\xspace}
\newcommand{\interworker}{inter-worker\xspace}

%% file: results_macros.tex
\newif\ifresnumcolor
\newcommand{\resnum}[1]{\ifresnumcolor\textcolor{red}{#1}\else#1\fi}

\newcommand{\rBgContextGap}{\resnum{29\ensuremath{\times}}\xspace}  %

\newcommand{\hbfCapacityRatio}{16\ensuremath{\times}\xspace}

\newcommand{\rTempLifetimeRatio}{\resnum{7.6--17.2\ensuremath{\times}}\xspace}  %

\newcommand{\rTempOverlap}{\resnum{0.96--0.99}\xspace}  %

\newcommand{\rTempByteFraction}{\resnum{66--83\%}\xspace}  %

\newcommand{\rStructLifetimeRatio}{\resnum{4.0--6.1\ensuremath{\times}}\xspace}  %

\newcommand{\rStructReviewerLifetimeRatio}{\resnum{4.7--38.3\ensuremath{\times}}\xspace}  %

\newcommand{\rStructOverlap}{\resnum{0.90--1.00}\xspace}  %

\newcommand{\rStructRwOverlap}{\resnum{0.25--0.60}\xspace}  %

\newcommand{\rStructLeadWaitFraction}{\resnum{71--88\%}\xspace}  %

\newcommand{\rInterWriteSpread}{\resnum{2.8--22.5\ensuremath{\times}}\xspace}  %

\newcommand{\rInterTtlSpread}{\resnum{3.7--23.6\ensuremath{\times}}\xspace}  %

\newcommand{\rInterSizeTtlCorr}{\resnum{0.51--0.72}\xspace}  %

\newcommand{\rInterSpawnKendall}{\resnum{0.53}\xspace}  %

\newcommand{\rInterBudgetCapture}{\resnum{62.8\%}\xspace}  %
\newcommand{\rInterBudgetWorst}{\resnum{none}\xspace}  %

\newcommand{\rEvalHbfLifetime}{\resnum{1.19--3.13\ensuremath{\times}}\xspace}  %

\newcommand{\rEvalWriteMixHbmBase}{\resnum{1--7\%}\xspace}        %
\newcommand{\rEvalWriteMixHbmLachesis}{\resnum{19--51\%}\xspace}  %
\newcommand{\rEvalWriteMixGain}{\resnum{8.2--19.5\ensuremath{\times}}\xspace}  %

\newcommand{\rEvalAblationRule}{\resnum{1.11--2.00\ensuremath{\times}}\xspace}  %
\newcommand{\rEvalAblationRank}{\resnum{1.33--1.54\ensuremath{\times}}\xspace}  %

\newcommand{\rEvalNoReviewer}{\resnum{1.05--1.14\ensuremath{\times}}\xspace}  %

\newcommand{\rEvalReviewerSpillRule}{\resnum{45--95\%}\xspace}      %
\newcommand{\rEvalReviewerSpillLachesis}{\resnum{2--25\%}\xspace}   %

\newcommand{\rEvalHbfAbsolute}{\resnum{3.3--12.2 device-years}\xspace}  %
\newcommand{\rEvalAbsMulti}{\resnum{3.3--9.3 device-years}\xspace}       %
\newcommand{\rEvalAbsMultiBase}{\resnum{1.9--4.8}\xspace}                %
\newcommand{\rEvalAbsSingle}{\resnum{4.6--12.2 device-years}\xspace}     %
\newcommand{\rEvalAbsSingleBase}{\resnum{2.1--6.4}\xspace}               %

\newcommand{\rEvalSweepGain}{\resnum{1.50--2.95\ensuremath{\times}}\xspace}  %

\newcommand{\rEvalSweepBase}{\resnum{12\%}\xspace}       %
\newcommand{\rEvalSweepLachesis}{\resnum{54--56\%}\xspace}   %

\newcommand{\rEvalLifetimeLoose}{\resnum{1.19--1.89\ensuremath{\times}}\xspace}  %

\newcommand{\rEvalHthreeTight}{\resnum{10.06--10.09\ensuremath{\times}}\xspace}  %
\newcommand{\rEvalHthreeLoose}{\resnum{2.53\ensuremath{\times}}\xspace}  %

\newcommand{\rEvalLifetimeSingle}{\resnum{1.35--3.13\ensuremath{\times}}\xspace}  %
\newcommand{\rEvalSingleAbsFall}{\resnum{30--31\%}\xspace}  %

\newcommand{\rDeployPeriod}{\resnum{five-year}\xspace}  %

\newcommand{\refTpot}{\resnum{0.2\,s}\xspace}  %
\newcommand{\refTtft}{\resnum{1.0\,s}\xspace}  %
\newcommand{\refLamUser}{\resnum{120\,s}\xspace}  %

\newcommand{\refLamTool}{\resnum{0.69\,s}\xspace}  %

\newcommand{\hbfPE}{\resnum{100{,}000}\xspace}  %

\newcommand{\hbfStackCap}{\resnum{512\,GB}\xspace}  %

\newcommand{\hbfWaf}{\resnum{1.02}\xspace}  %

\newcommand{\sloTightTpot}{\resnum{60\,ms}\xspace}
\newcommand{\sloLooseTpot}{\resnum{100\,ms}\xspace}
\newcommand{\sloTtft}{\resnum{2\,s}\xspace}

%% file: contents/introduction.tex
\section{Introduction}
\label{sec:introduction}

\begin{figure*}[t]
  \centering
  \includegraphics[page=1,width=\textwidth]{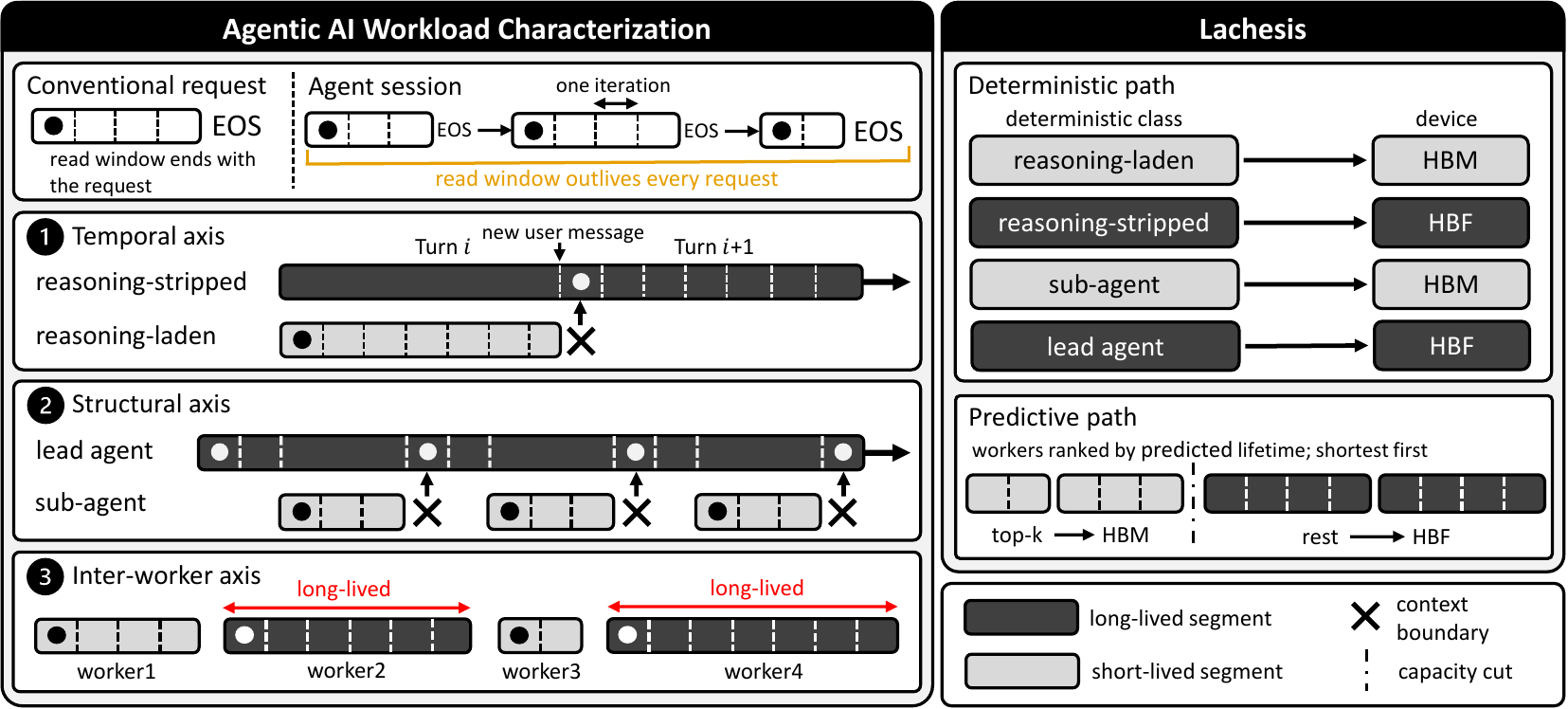}
  \caption{Top left: the read window of a conventional request, and of an agent session.
  Left: the three axes along which context-segment lifetimes differ.
  Right: \lachesis's two placement paths.}
  \label{fig:taxonomy}
  \Description{A schematic in two halves. The left half opens on a contrast: a conventional request, whose read window ends where the request ends, beside an agent session of three requests, each continuing into the next at its EOS, spanned by one read window. Below it, one agent session's context is drawn on three axes: a turn end, a sub-agent return, and a strip of worker segments that differ in lifetime. The right half names a device for each segment class the harness fixes, sending reasoning-laden and the sub-agent segment to HBM, and reasoning-stripped and the lead-agent segment to HBF. Where the class is the same for every worker, it ranks worker invocations by predicted lifetime, shortest first, and a dashed capacity cut sends the top-$k$ to HBM and the rest to HBF. A legend gives the fill for a long-lived segment and for a short-lived segment, the cross for a context boundary, and the dashed rule for the capacity cut.}
\end{figure*}

LLM serving is shifting toward agentic workloads, in which an agent works on a single user task over many LLM requests and spawns sub-agents that run concurrently alongside it~\cite{yuan2026agentic}.
Whereas a conventional request holds one context for the duration of one request, an agent holds the context it has accumulated so far in the session as KV cache, and that KV cache can stay live well beyond the request that wrote it, in some cases until the session ends (Figure~\ref{fig:taxonomy}, top left).
Together with the sub-agents a session spawns, this longer lifetime pushes the volume of KV cache live at one time far beyond that of conventional serving.
The pressure this KV cache puts on an accelerator's high-bandwidth memory (HBM) forces a deployment to add accelerators or serve fewer sessions at once, either of which raises the hardware cost of a served request~\cite{ha2026h3,kyung2026hbfkv,son2026hbf}.

High-bandwidth flash (HBF) is emerging as a promising class of memory for LLM inference, offering an order of magnitude greater capacity than HBM with HBM-class sequential read bandwidth~\cite{sandisk2025hbf}.
That capacity absorbs the added pressure and brings the cost of a served request back down.
The key limiting factor for HBF, however, is write endurance.
NAND flash tolerates only a finite number of program/erase (P/E) cycles, so a DRAM-like role that absorbs a large volume of writes can wear it out in a fraction of HBF's required service life.
Early HBF designs therefore confine HBF to read-only data such as model weights and shared-prefix KV caches~\cite{ha2026h3,oliveira2026flint,kim2026beyondcapacity}.
As KV caches are a dominant source of serving-time memory pressure in agentic and large-batch workloads~\cite{liu2024kivi,qin2026mooncake,liu2025lmcache}, recent works extend HBF to dynamically generated KV cache and select what to write to it, mainly by its read-to-write ratio~\cite{kyung2026hbfkv,wang2026flashaccel,park2026hbm}.

Unlike prior work, which selects KV cache for HBF by its read-to-write ratio~\cite{kyung2026hbfkv}, we select it by its lifetime. When HBF is used alongside HBM, every KV block a session writes lands in one of the two tiers, and only the bytes that land in HBF spend endurance. HBF therefore lasts longer the more of a session's writes HBM absorbs.
The volume HBM absorbs over a period is its capacity multiplied by the number of times each block is reclaimed when the KV cache it holds is discarded. Placing shorter-lived KV blocks in HBM therefore lets HBM absorb more of a session's writes and sends less of them to HBF, which requires knowing a block's lifetime before it is written.

In agentic serving, KV lifetime is set by the agent harness, the program that orchestrates the session~\cite{ruan2026aorchestra,wang2025openhandssdk,anthropic2026subagents,langchain2026deepagents}, so we analyze how the harness manages it. Depending on how it structures the session and the sub-agents it spawns, the harness keeps some KV cache for the whole session and discards the rest partway through. It makes that decision for KV in groups, so each group shares one lifetime. We call such a group a \emph{context segment} and the point at which it is discarded a \emph{context boundary}.

Our analysis shows that segment lifetimes diverge along three axes of session structure, shown in the left panel of Figure~\ref{fig:taxonomy}.
The \emph{temporal} axis follows the \emph{turns} of a session, one opened by each user message. When the next message arrives, the harness strips the reasoning generated within the turn and carries only its output forward. The reasoning segment holds \rTempByteFraction of the KV the turn writes yet lives only one turn, whereas the output segment lives \rTempLifetimeRatio longer.
The \emph{structural} axis follows delegation within a turn. A \emph{lead agent} may, depending on the harness, delegate subtasks to \emph{sub-agents}, each running in its own context. When a sub-agent returns, the harness discards its segment and keeps only its result. Sub-agent segments produce most of the KV written within the turn, while the lead-agent segment lives \rStructLifetimeRatio longer.
The temporal and structural boundaries are both predetermined and known to the harness. On those two axes the short-lived segment, which takes large portion of writes, can therefore be placed in HBM at the write and freed the moment the boundary arrives.
The third axis, \interworker, follows the \emph{workers} that carry out those subtasks, whose lifetimes differ by when each completes. That completion time is not fixed in advance, and worker lifetimes spread by \rInterTtlSpread between p10 and p90, so segments of one class carry significantly different lifetimes. Our key observation is that their relative order can be predicted at spawn time from the record the harness uses to build each worker's context. Placement can thus be decided before the KV is written along all three axes.

We present \lachesis, a lifetime-aware KV cache placement layer that decides which KV reaches HBF before a segment is written (\secref{sec:placement}).
\lachesis places along two paths (Figure~\ref{fig:taxonomy}, right).
The \emph{deterministic path} covers the temporal and structural axes, where the harness knows whether a segment is discarded at the boundary or retained past it. A \shortlived has priority for HBM, and a \longlived is placed in HBF.
The \emph{predictive path} covers the \interworker axis, where no context boundary separates one worker's segment from another's. \lachesis predicts each segment's lifetime at spawn, ranks the segments shortest-first, and admits as many as HBM has room for, placing the rest in HBF.
Lifetime is tracked entirely in \lachesis, and the serving engine sees it only through two calls.
The first allocates a KV block from the tier \lachesis specifies, HBM or HBF, so every placement decision of either path is issued as one allocation.
The second frees the blocks of a segment whose lifetime has ended, at the boundary the harness reports.
Across agent traces replayed through a simulated two-tier memory system, \lachesis extends HBF lifetime by \rEvalHbfLifetime over HBM-first placement, which allocates in HBM while it has a free block and in HBF otherwise (\secref{sec:evaluation}).
At a tight service-level objective (SLO) under continuous 24$\times$7 operation, HBF outlasts a \rDeployPeriod warranty on the multi-agent trace.

In summary, this paper makes the following contributions:

\begin{itemize}
  \item We show that the context an agent session accumulates divides into segments of differing lifetime along three axes, temporal, structural, and \interworker, and that a segment's lifetime can be determined before it is written.
  \item We design \lachesis, a lifetime-aware KV cache placement layer between the agent harness and the serving engine. Its deterministic path places the segments the harness marks as short-lived in HBM, and its predictive path admits the workers predicted to be shortest-lived, so that less KV reaches HBF and its endurance lasts longer.
  \item We evaluate \lachesis against HBM-first placement and show that it extends HBF lifetime by \rEvalHbfLifetime, and that HBF then outlasts a \rDeployPeriod warranty on the multi-agent trace at a tight SLO.
  A session writes the same KV volume under either policy; \lachesis admits the shortest-lived segments to HBM and frees each at its boundary, so the same HBM capacity absorbs more of that volume and less of it reaches flash.
\end{itemize}
\phantomsection\label{paper:introend}

%% file: contents/background.tex
\section{Background and Motivation}
\label{sec:background}

\subsection{Agent Serving Intensifies the Capacity Wall}
\label{sec:bg-capacity}

Agentic serving is increasingly constrained by a \emph{memory-capacity wall}, pushing up the cost of serving each request.
The pressure grows from two sides, the model footprint and the KV cache.
The model footprint has grown faster than per-accelerator HBM capacity, requiring more accelerators simply to host a model \cite{ha2026h3,son2026hbf}.
The other side is the KV cache, which grows with the context each session accumulates.

\begin{figure}
  \centering
  \begin{subfigure}{\columnwidth}
    \centering
    \includegraphics[page=3,width=0.9\linewidth]{figures/figures_crop}
    \caption{Execution structure of an agent session, single-agent above and multi-agent below.}
  \end{subfigure}
  \par\medskip
  \begin{subfigure}{\columnwidth}
    \centering
    \includegraphics[trim=0 13.5bp 0 120bp,clip,width=\linewidth]{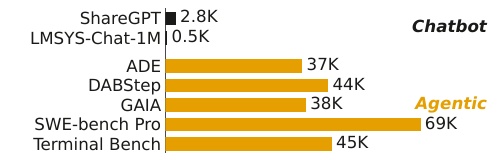}
    \caption{Mean peak context per conversation or agent task.}
  \end{subfigure}
  \caption{Context accumulation in agent serving.}
  \label{fig:agent-context}
  \Description{Two stacked panels. Panel (a) is a schematic in two halves, single-agent above and multi-agent below. Its upper half has two rows, a multi-turn session of a user message followed by multi-round reasoning followed by the next user message, and an expansion of one multi-round block into successive rounds, each a reasoning step, a tool call and an observation, ending in an output. Its lower half also has two rows, the lead agent, whose multi-round reasoning branches down at a sub-agent spawn and resumes after it, and the sub-agent, which receives a lead-agent message, performs its own multi-round reasoning, and returns into the lead agent's next reasoning block. Panel (b) is a horizontal bar chart of mean peak context, with two chatbot corpora at the top and five agentic workloads below, the agentic bars far longer than the chatbot bars.}
\end{figure}

An agent session accumulates context by repeating the reason-act-observe loop of a ReAct-style agent until the task is done (Figure~\ref{fig:agent-context}(a), top).
Each pass is a \emph{round}, a single LLM request, and it appends all three to the context every later round must read \cite{yao2023react,yuan2026agentic}.
A session divides into \emph{turns}, one per user message, and a turn holds one or more rounds.
An observation is sized by the environment rather than by a person, so a source file or a command's output enters the context whole.
The peak context lengths of agentic workloads are \rBgContextGap larger than those of conventional chatbot workloads (Figure~\ref{fig:agent-context}(b)).

\begin{figure}
  \centering
  \includegraphics[width=\columnwidth]{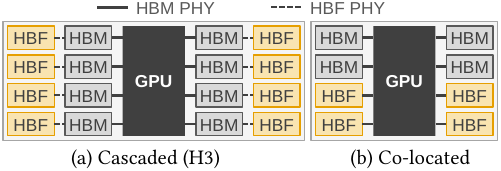}
  \caption{GPU architectures equipped with HBF.}
  \label{fig:hbf-arch}
  \Description{Two block diagrams of an accelerator package, each with a GPU die at the center and memory stacks on both sides. A solid line marks the HBM PHY and a dashed line the HBF PHY. In (a), cascaded, every row is an HBM stack with an HBF stack chained beyond it, so flash reaches the GPU through the HBM interface. In (b), co-located, the upper rows are HBM stacks and the lower rows HBF stacks, each attached to the GPU by its own PHY.}
\end{figure}

A multi-agent session holds more than one such context at a time (Figure~\ref{fig:agent-context}(a), bottom).
A \emph{lead agent} carries the user's task and delegates a subtask to a \emph{sub-agent}, which runs its own rounds in a context of its own and returns a result \cite{anthropic2026multiagent,anthropic2026subagents,kimi2026k25}.
The session therefore occupies the lead agent's context and one more for every sub-agent running under it.
Many such sessions are served at once, and each holds its accumulated context in HBM.
Deployments must either spread each request across more accelerators or reduce the number of concurrently resident contexts, both of which increase the hardware cost per served request \cite{ha2026h3,son2026hbf,kyung2026hbfkv}.

Existing KV-cache optimizations reduce memory demand, but do not eliminate the capacity wall.
Quantization and compression reduce the bytes per cached token only by a constant factor, while KV capacity still grows linearly with context length \cite{liu2024kivi,nawrot2024dmc}.
Eviction and a bounded attention window cap the KV footprint by discarding past KV, which degrades accuracy on long-context workloads \cite{zhang2023h2o,xiao2024streamingllm}.
Sparse attention reduces the KV read per step, but the KV footprint remains unchanged \cite{tang2024quest}.
Thus, serving long agentic contexts still exceeds HBM capacity, demanding a substantially larger memory tier.

\subsection{High-Bandwidth Flash: Capacity Opportunity, Endurance Constraint}
\label{sec:bg-hbf}

\input{tables/t2_axes}
\input{tables/t1_regimes}

HBF, an emerging stacked memory built from NAND flash, can relieve the memory-capacity wall while providing HBM-class sequential read bandwidth.
HBF is projected to provide \hbfCapacityRatio the per-stack capacity of HBM, and to exploit thousands of NAND planes to deliver high sequential bandwidth \cite{sandisk2025hbf, ocp_hbf}.
Its larger capacity allows more contexts to remain resident on each accelerator, while its read bandwidth is sufficient to serve attention efficiently.

HBF is therefore a promising alternative to HBM, and Figure~\ref{fig:hbf-arch} shows two architectures explored by prior works that directly integrate HBF with accelerators.
Daisy-chained designs connect HBF stacks behind HBM, allowing HBF traffic to reach the accelerator through the HBM stacks (Figure~\ref{fig:hbf-arch}(a)) \cite{ha2026h3}.
Co-located designs instead attach HBF directly alongside the accelerator (Figure~\ref{fig:hbf-arch}(b)) \cite{son2026hbf,kyung2026hbfkv}.
When accessing HBF-resident data in co-located designs, the accelerator can utilize the aggregate memory bandwidth of all stacks through direct and inter-stack paths \cite{kim2026beyondcapacity, an2026memory}.

Prior works have addressed the main performance limitations of HBF, which are outside the scope of this work \cite{wang2026flashaccel, ha2026h3, oliveira2026flint, kim2026beyondcapacity}.
First, HBF read latency is orders of magnitude higher than HBM.
LLM inference exposes predictable sequential accesses that enable aggressive prefetching to hide this latency.
Second, sustaining HBF's peak read bandwidth requires accessing many flash planes in parallel; prior work achieves this through plane-balanced data layouts.

The remaining obstacle to using HBF in LLM serving is the limited write endurance.
Because NAND supports only a finite number of P/E cycles, prior HBF designs have largely focused on read-only state that is written once and then repeatedly read \cite{sandisk2025hbf}.
Designs such as H3 place model weights and precomputed KV cache in HBF while retaining online-generated KV in HBM \cite{ha2026h3,oliveira2026flint,kim2026beyondcapacity}.

For agentic serving, however, the generated KV becomes the dominant capacity bottleneck, making write endurance a first-order challenge.
Unlike model weights, whose footprint is fixed once the model is loaded, generated KV grows with context length and the number of resident sessions (Figure~\ref{fig:agent-context}(b)).
Fully exploiting HBF, therefore, requires placing the generated KV in HBF \cite{son2026hbf,kyung2026hbfkv}.
The KV cache is generally well suited to NAND flash.
Each entry is written once during prefill or decode and then repeatedly read by subsequent attention steps \cite{kwon2023vllm}.
But short-lived KV with little reuse can nonetheless quickly exhaust write endurance~\cite{son2026hbf}.

\subsection{Why Agent Sessions Require Lifetime-Aware KV Cache Placement}
\label{sec:bg-why-lifetime}

\begin{figure*}
  \centering
  \includegraphics[width=\textwidth]{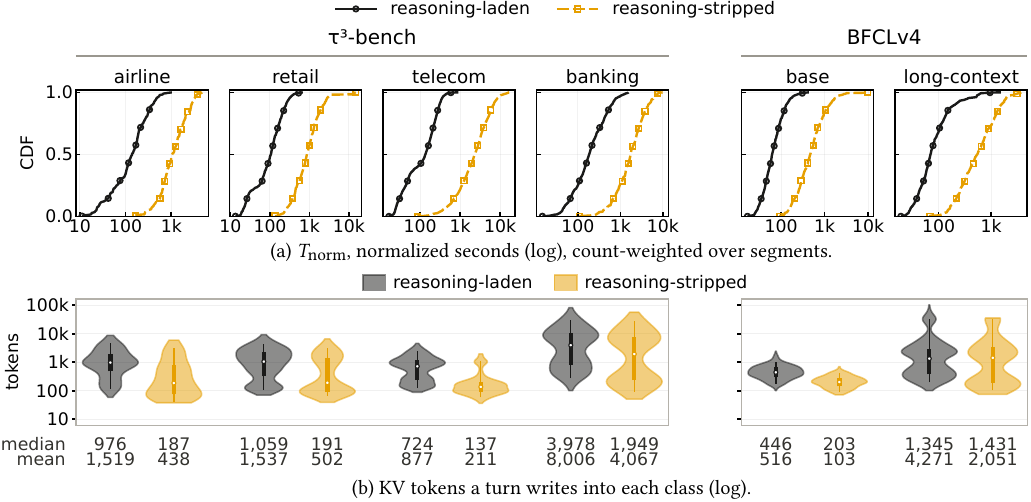}
  \caption{Temporal axis, per domain.
  $\tnormbare$ is drawn at one sampled user delay, median \refLamUser, and one sampled tool-call latency, mean \refLamTool.}
  \label{fig:temporal-distribution}
  \Description{Two rows of plots over the same six domains in the same order. Row (a) holds one panel per domain, plotting cumulative distributions of normalized context lifetime for reasoning-laden and reasoning-stripped context segments, one point per segment; in every panel the reasoning-stripped curve lies to the right of the reasoning-laden curve. Row (b) holds one plot area per benchmark group, four domains in the first and two in the second, each group sharing one vertical axis. Each area carries a pair of violins of the KV tokens a turn writes into each class at every domain position, reasoning-laden first and reasoning-stripped second, with the median and mean printed beneath each violin and a key to the pair centred above the row.}
\end{figure*}

Although Kyung et al. propose placing the KV cache with a high read-to-write ratio in HBF \cite{kyung2026hbfkv}, this criterion does not directly apply to agentic serving.
They show that output length largely determines the read-to-write ratio because longer generation reuses the KV cache across more decode steps.
This relationship holds for stateless requests, where all KV is created and consumed within a single request and therefore shares the request's lifetime.
Agent sessions are stateful, however, and different parts of their context can remain useful for different lengths of time.
The harness may discard some context while retaining the rest for later turns or subtasks.
We call a group of tokens that share the same lifetime a \emph{context segment}.

HBF placement should therefore be decided per context segment rather than per request or session.
Keeping short-lived segments in HBM reduces the write traffic reaching flash, thereby extending HBF lifetime.
In the following section, we analyze agentic serving workloads to identify the sources of lifetime variation and characterize which context segments are short- or long-lived.

%% file: tables/t2_axes.tex
\begin{table*}
  \centering
  \small
  \caption{The three characterization axes. The labels name the two sides of each axis's own comparison, not one threshold across axes.}
  \label{tab:axes}
  \begin{tabularx}{\textwidth}{L{0.82}L{0.88}L{0.90}L{1.00}L{1.42}L{0.98}}
    \toprule
    Axis & Scope relationship & What creates the split & Short-lived & Long-lived & Signal \\
    \midrule
    Temporal (\S\ref{sec:ana-temporal})
      & session $\supset$ turn
      & Reasoning stripping
      & \ladenseg
      & \strippedseg
      & Deterministic boundary \\
    \addlinespace
    Structural (\S\ref{sec:ana-structural})
      & session $\supset$ sub-agent
      & Sub-agent return
      & \subseg
      & \leadseg
      & Deterministic boundary \\
    \addlinespace
    Inter-worker (\S\ref{sec:ana-interworker})
      & across worker invocations
      & Variation in return time
      & Early-returning worker's segment
      & Late-returning worker's segment
      & Predicted rank \\
    \bottomrule
  \end{tabularx}
\end{table*}

%% file: tables/t1_regimes.tex
\begin{table*}
  \centering
  \small
  \caption{Execution regimes. Each is configured so that exactly one context boundary is active. Both regimes use Gemma~4.}
  \label{tab:regimes}
  \begin{tabularx}{\textwidth}{L{0.48}L{1.18}L{2.05}L{0.92}L{0.87}L{0.50}}
    \toprule
    Regime & Harness & Benchmarks & Session shape & Active boundary & Axes \\
    \midrule
    Temporal
      & Harbor, OpenAI message format, model's own chat template
      & $\tau^3$-bench~\cite{yao2024taubench,sierra2026tau3bench}, BFCLv4 multi-turn~\cite{patil2025bfcl}
      & Multi-turn, single agent
      & Turn-end reasoning stripping
      & \S\ref{sec:ana-temporal} \\
    \addlinespace
    Structural
      & AOrchestra (inherits Claude Code)
      & AppWorld~\cite{trivedi2024appworld}, DeepSearchQA~\cite{gupta2026deepsearchqa}, SWE-bench Verified~\cite{jimenez2024swebench}, TerminalBench~\cite{merrill2026terminalbench}, WebArena~\cite{zhou2024webarena}
      & Single-turn, multi-round lead-agent/sub-agent
      & Sub-agent return
      & \S\ref{sec:ana-structural}, \S\ref{sec:ana-interworker} \\
    \bottomrule
  \end{tabularx}
\end{table*}

%% file: contents/analysis.tex
\section{Context-Lifetime Analysis for Agent Serving}
\label{sec:analysis}

\subsection{Analysis Overview}
\label{sec:ana-framework}

Table~\ref{tab:axes} summarizes the three characterization axes that arise in agentic serving, where each axis separates short-lived from long-lived context segments.
Along the temporal axis, reasoning context is discarded at the turn end, and the remaining reasoning-stripped context merges with the session context (\secref{sec:ana-temporal}).
Similarly, along the structural axis, sub-agent context excluding the result is discarded at the sub-agent return (\secref{sec:ana-structural}).
The discarded segment is a \shortlived, and what survives the boundary is a \longlived.
Along the inter-worker axis, the class is constant and worker segments differ in lifetime by when each worker returns (\secref{sec:ana-interworker}).

The harness fixes a segment's class before the segment is written, so the lifetime is known at the write.
The first two splits are deterministic, fixed at the turn end and at sub-agent spawn.
The last split, worker lifetime, is predictable from the metadata the harness records when it spawns each worker.
A reactive policy such as least-recently-used (LRU) cannot assign the class before observing subsequent accesses and would require migration to correct the placement later.

\subsection{Methodology}
\label{sec:methodology}

\begin{figure*}
  \centering
  \includegraphics[width=\textwidth]{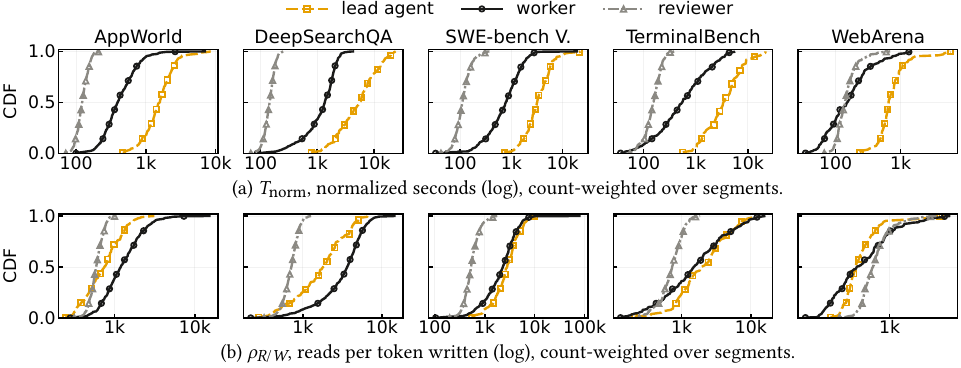}
  \caption{Structural axis, per benchmark.
  $\tnormbare$ is drawn at one sampled tool-call latency per tool class.}
  \label{fig:structural-distribution}
  \Description{A grid of panels in two rows, one column per benchmark. Row (a) plots cumulative distributions of normalized context lifetime for lead-agent, worker and reviewer segments; row (b) plots cumulative distributions of the read-to-write ratio for the same three series. In row (a) the lead-agent curve lies to the right of the two sub-agent series.}
\end{figure*}

We isolate the turn end and sub-agent return in two execution regimes, each configured so that only the context boundary under study can close a read window (Table~\ref{tab:regimes}).
A lifetime difference measured in a regime is therefore attributable to that regime's boundary.
The temporal regime runs on Harbor \cite{harbor2026framework} and the structural regime on AOrchestra, which builds each sub-agent from an explicit $\langle$Instruction, Context, Tools, Model$\rangle$ specification at spawn \cite{ruan2026aorchestra}.
We call that specification the \emph{spawn record}, from which \secref{sec:ana-interworker} ranks worker lifetimes.
No harness records segments, so they are recovered from a per-request trace.
The trace records the exactly matched input-prefix length and the cache read SGLang reports \cite{zheng2024sglang}.
A segment's read window closes where the segment stops appearing in the context that later requests reuse, or at the end of the session if no boundary closes it.
Rematerialization produces a new segment, not an in-place rewrite, so each segment is materialized once with its own write volume.

The primary metric, normalized context lifetime, is how long a segment must remain readable on a fixed reference clock rather than on server wall time:
\begin{equation}
  \tnormeq
  \label{eq:tnorm}
\end{equation}
Every term is counted while segment $\seg$ is live: $\Gdec$ is the number of tokens generated on the decode node, $\Nobs$ the number of observation prefills, $\lamtool$ the tool-call latency, and $\lamuser$ the user delay before the next message.
The clock is the decode node's clock in a prefill--decode disaggregated deployment \cite{zhong2024distserve,patel2024splitwise,feng2025windserve}.
Its reference point is a time per output token $\tpotref$ = \refTpot \cite{jin2023s3,kakolyris2025throttllem,cho2026pascal} and a time to first token $\ttftref$ = \refTtft \cite{cheng2026kunserve,zhang2025blitzscale}.
Normalizing removes decode batching, server load, and queueing delay, which are properties of the serving deployment rather than of the agent session.
$\lamtool$ and $\lamuser$ are modeled rather than taken from the trace, because tool-call latency and user delay depend on the environment an agent is deployed in (\secref{sec:eval-setup}).
The read-to-write ratio $\rhorw$, by which prior work selects KV for HBF \cite{kyung2026hbfkv}, divides the KV read from a segment by the KV written to materialize it.
Each session is analyzed independently, and no measurement aggregates across concurrent sessions.

\subsection{Temporal Axis: Reasoning-Content Stripping}
\label{sec:ana-temporal}

Along the temporal axis, reasoning determines how long a context segment remains live.
Across multiple rounds within a turn, the harness accumulates reasoning context across tool calls along with non-reasoning context \cite{openai2026reasoning,deepseek2026thinking}.
The accumulated context is read throughout the later rounds in the same turn \cite{yuan2026agentic}.
When the turn ends as the next user message arrives, the harness discards the reasoning context, the \ladenseg, and only the remaining \strippedseg from the previous turn is passed to the new turn \cite{openai2026reasoningitems}.

Figure~\ref{fig:temporal-distribution}(a) shows that the \strippedseg outlives the \ladenseg by a median ratio of \rTempLifetimeRatio across all domains.
The \strippedseg is retained across multiple consecutive turns, while the \ladenseg is retained only during the single turn that generates the context.
The lifetime distributions of the two segments are clearly separated: a separability, the probability that a random \strippedseg outlives a random \ladenseg, is \rTempOverlap.

Placing short-lived \ladenseg in HBF is particularly costly because it is much larger than the \strippedseg.
Of the tokens written by a turn, \rTempByteFraction belong to the \ladenseg (Figure~\ref{fig:temporal-distribution}(b)).
These tokens consume extensive HBF P/E cycles even though the harness soon discards them at the next turn end.
The same write budget can instead retain many smaller, long-lived contexts in HBF.

\begin{takeaway}{1}
\setlength{\leftmargini}{1em}\setlength{\labelsep}{0.35em}
\begin{itemize}\itemsep0pt\parsep0pt
\item The \ladenseg is short-lived yet holds \rTempByteFraction of the tokens a turn writes, while the \strippedseg that replaces it is long-lived.
\item HBM should hold the \shortlived and HBF the \longlived.
\end{itemize}
\end{takeaway}

\subsection{Structural Axis: Lead-Agent/Sub-Agent Isolation}
\label{sec:ana-structural}

In multi-agent sessions, the lead-agent context is retained longer than the context created for each delegated subtask.
The lead agent delegates a subtask to a sub-agent, which executes the task in a separate context.
When the sub-agent returns, only its result is merged into the \leadseg, while the remaining \subseg is discarded.
This separation is by design of the multi-agent execution model, not an optimization specific to one harness \cite{anthropic2026multiagent,openai2026agentsastools,anthropic2026subagents,langchain2026deepagents,kimi2026k25}.
Therefore, \leadseg contains session-level context, including the system prompt, user objective, spawn records, and returned sub-agent results, and \subseg contains the context of a single sub-agent, including its prompt, spawn instruction, and generated content.
While \leadseg remains available until the session ends, \subseg is used only until that sub-agent returns.

The \subseg is consistently shorter-lived than the \leadseg across all benchmarks.
The lifetime of \subseg varies by role: a worker actively executes a delegated task through multiple rounds of tool use, while a reviewer receives a completed worker trace and produces a review without further delegation.
Figure~\ref{fig:structural-distribution}(a) shows that the \leadseg outlives a worker's segment by a median ratio of \rStructLifetimeRatio and a reviewer's by \rStructReviewerLifetimeRatio, with separability \rStructOverlap between their lifetime distributions.

For HBF, placing the long-lived \leadseg is beneficial although its reuse is comparable to that of a worker sub-agent.
After the lead agent delegates the subtasks, sub-agent decoding, observation prefills, and tool-call latency increase the lead agent's $\tnormbare$ while the \leadseg remains unread.
This waiting period accounts for \rStructLeadWaitFraction of the lead-agent lifetime.
As a result, the lead agent and worker overlap in $\rhorwbare$ by \rStructRwOverlap despite their large difference in lifetime (Figure~\ref{fig:structural-distribution}(b)).
Reuse should therefore not be used as a proxy for lifetime when deciding HBF placement.

\begin{takeaway}{2}
\setlength{\leftmargini}{1em}\setlength{\labelsep}{0.35em}
\begin{itemize}\itemsep0pt\parsep0pt
\item The \subseg is short-lived, while the \leadseg is long-lived yet unread for \rStructLeadWaitFraction of that lifetime.
\item HBM should hold the \subseg and HBF the \leadseg.
\end{itemize}
\end{takeaway}

\subsection{Inter-Worker Axis: Lifetime Heterogeneity and Spawn-Time Predictability}
\label{sec:ana-interworker}

\begin{figure}
  \centering
  \includegraphics[width=\columnwidth]{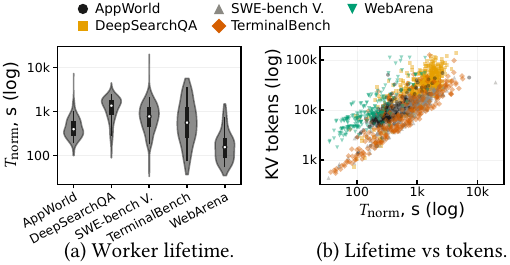}
  \caption{Worker heterogeneity, per benchmark.}
  \label{fig:interworker-heterogeneity}
  \Description{Two panels side by side, under a shared legend of five benchmarks. Panel (a), worker lifetime, holds one violin per benchmark against a logarithmic axis of normalized context lifetime in seconds, each violin marked with its median and interquartile box. Panel (b), lifetime against tokens, is a scatter of one point per worker segment, normalized context lifetime on the horizontal axis and the KV tokens that segment writes on the vertical, both logarithmic, with a marker and colour per benchmark.}
\end{figure}

All worker segments belong to the sub-agent class, yet their lifetimes vary widely, and no context boundary separates them.
Within a single benchmark, worker $\tnormbare$ varies by \rInterTtlSpread from p10 to p90 (Figure~\ref{fig:interworker-heterogeneity}(a)), and worker write volume varies by \rInterWriteSpread over the same percentiles.
The two correlate at Kendall's $\tau$ of \rInterSizeTtlCorr per benchmark (Figure~\ref{fig:interworker-heterogeneity}(b)), so a worker that writes more usually holds a longer-lived segment.
A session runs many sub-agents, so the sub-agent class accounts for most of what a session writes, and treating it as homogeneous ignores that within-class spread.

Lifetime and write volume both scale with a worker's round count, but only lifetime also includes tool-call latency.
A worker's $\tnormbare$ is a round count times a per-round cost, which Equation~\ref{eq:tnorm} splits into a generation-side term, $\Gdec\tpotref + \Nobs\ttftref$, and a tool-call latency term, $\lamtool$.
Write volume grows with the same round count, so the two spreads share that factor.
Tool-call latency adds no tokens, so a waiting worker advances its $\tnormbare$ without adding to its write volume.
This latency differs greatly from tool to tool, so workers that write similar volumes can still have very different lifetimes.
Published measurements put mean call durations three to four orders of magnitude apart, and tool execution takes up to nearly a third of end-to-end agent latency \cite{yuan2026agentic}.

\begin{figure}
  \centering
  \includegraphics[width=\columnwidth]{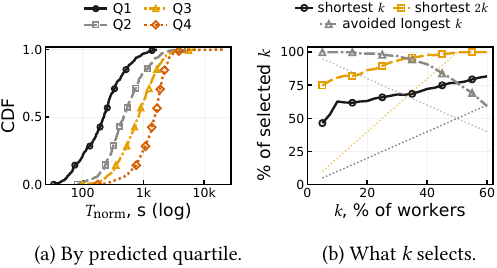}
  \caption{Spawn-time lifetime prediction.}
  \label{fig:interworker-ttl-cdf}
  \Description{Two panels side by side. Panel (a), by predicted quartile, plots four cumulative distributions of measured worker normalized context lifetime against a logarithmic horizontal axis, one curve per quartile of the predicted ranking, Q1 shortest predicted to Q4 longest, and the four lie in that order from left to right. Panel (b), what $k$ selects, plots three shares of the selected $k$ against $k$ as a percentage of workers taken shortest-first. They are the share falling among the true shortest-lived $k$, the share falling among the true shortest-lived $2k$, and the share that avoided the true longest-lived $k$, each beside its own dotted chance level.}
\end{figure}

A worker's lifetime and write volume are known only after it finishes, so we predict its lifetime rank at spawn.
The target is the rank rather than the value, because estimating $\tnormbare$ itself is hard while recovering the order is not.
Analysis of a stateless request identifies that request's output length as the signal \cite{kyung2026hbfkv}; the quantity ranked here is the worker's whole lifetime.
The predictor reads the spawn record's instruction, context and tool list, together with where the spawn falls in the session; we call that set the worker's \emph{spawn signal}.
The predictor is a linear model over that signal, fitted by ridge regression and read only for the order it induces.
Each element enters the model as a length, an indicator per tool, a session position, or a hashed bag of its words.
A single predictor serves all five benchmarks.

The predicted ranking orders workers by lifetime at Kendall's $\tau$ of \rInterSpawnKendall, a metric that scores the order alone.
The predictor is scored by five-fold cross-validation that keeps all workers of a session in the same fold, so no worker is scored against training data from its own session.
Figure~\ref{fig:interworker-ttl-cdf}(a) groups worker $\tnormbare$ by predicted rank, from Q1 shortest to Q4 longest, and the four quartiles separate across the whole range.
Figure~\ref{fig:interworker-ttl-cdf}(b) takes the $k$ shortest-predicted workers at each $k$ and compares three shares of that selection with a random selection of the same size, the share among the $k$ shortest-lived, the share among the $2k$ shortest-lived, and the share that avoids the $k$ longest-lived.
At one worker in ten, the selection holds \rInterBudgetCapture of the shortest-lived tenth and \rInterBudgetWorst of the longest-lived tenth.

\begin{takeaway}{3}
\setlength{\leftmargini}{1em}\setlength{\labelsep}{0.35em}
\begin{itemize}\itemsep0pt\parsep0pt
\item Within the worker role, lifetime varies by \rInterTtlSpread from p10 to p90, and no context boundary separates that spread.
\item Workers are ranked by a lifetime predicted at spawn, and the segments ranked first have priority for HBM.
\end{itemize}
\end{takeaway}

On the temporal and structural axes, the segment class the harness fixes is available when the segment is written.
Among workers, where no context boundary separates them, the spawn record exposes candidate signals for the variation inside that one class.

%% file: contents/placement.tex
\section{\texorpdfstring{\lachesis}{Lachesis} Design}
\label{sec:placement}

\begin{figure}
  \centering
  \includegraphics[page=2,width=\columnwidth]{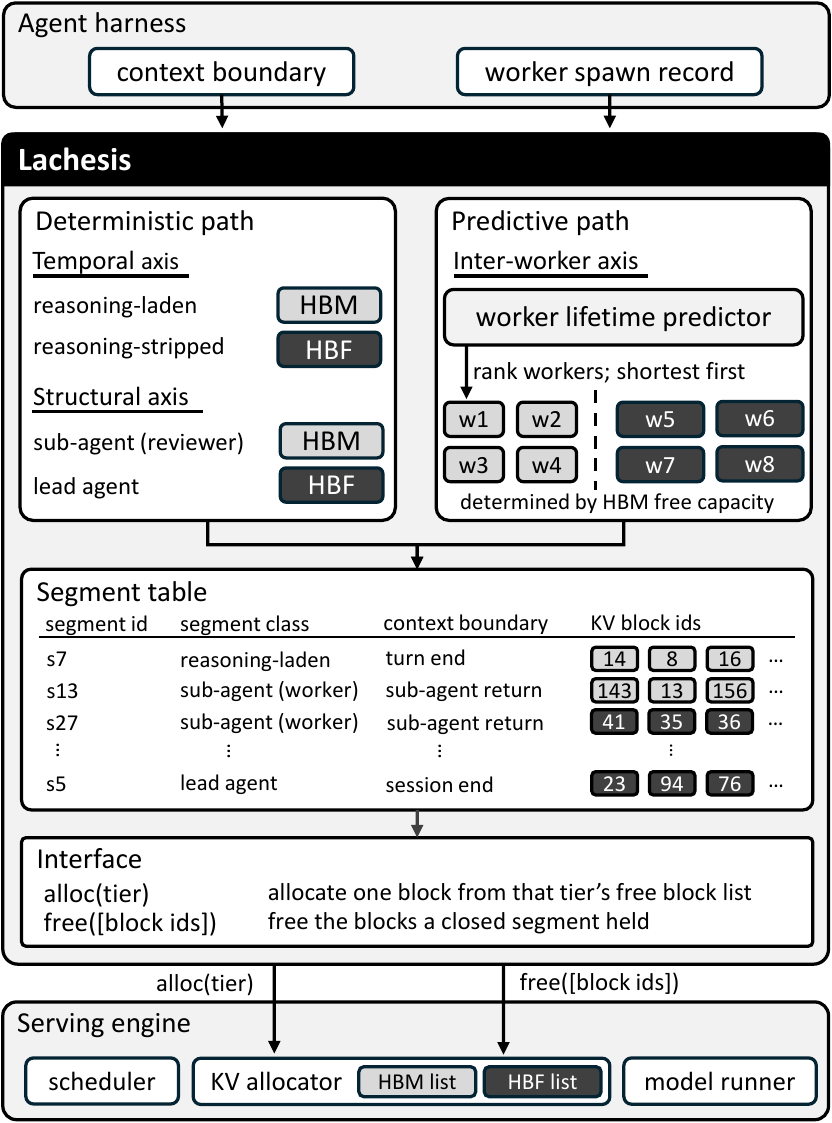}
  \caption{Overview of \lachesis.}
  \label{fig:placement}
  \Description{A block diagram. An agent harness lies above and a serving engine below, with \lachesis between them. Two arrows run down from the harness into \lachesis, one for context boundaries and one for the worker spawn record. \lachesis holds two paths side by side, and their outputs join into one arrow running down to the blocks beneath them. The deterministic path carries the temporal and structural axes, mapping reasoning-laden and the reviewer invocation to HBM, and reasoning-stripped and the lead-agent segment to HBF. The predictive path carries the \interworker axis, ranking worker invocations by predicted lifetime, shortest first, and admitting as many as HBM's free capacity determines. Below them a segment table holds one row per live segment, keyed by segment id and giving the segment class, the context boundary that closes it, and the KV blocks it holds. Below the table an interface block names the two calls, one allocating a block from a named tier's free block list and one freeing the blocks a closed segment held. Both calls run downward out of \lachesis into the engine's KV allocator, which keeps an HBM free list and an HBF free list.}
\end{figure}

Based on the three axes of \secref{sec:analysis}, we present \lachesis, a lifetime-aware KV cache placement layer that decides which KV reaches HBF before a segment is written.
As Figure~\ref{fig:placement} shows, \lachesis lies between an agent harness such as the OpenHands SDK, AOrchestra or Claude Code \cite{wang2025openhandssdk,ruan2026aorchestra,anthropic2026subagents} and a serving engine such as vLLM or SGLang \cite{kwon2023vllm,zheng2024sglang}.
\lachesis receives three things from the harness, the id of the segment that each LLM request extends, the spawn record written at each sub-agent spawn, and the context boundary that closes a segment.
It holds three components, a deterministic path that reads the segment class that the harness fixes on the temporal and structural axes, a predictive path that ranks workers by a lifetime predicted at spawn on the \interworker axis, and a segment table that carries one row for each live segment.
It reaches the serving engine as two calls, one that allocates a block in a named tier and one that frees the blocks of a closed segment.
Its placement covers KV only, leaving \secref{sec:bg-hbf}'s read-only mapping for model weights unchanged and activations in HBM.

The deterministic path places the \shortlived in HBM and the \longlived in HBF, by the segment class that the harness fixes.
In a single-agent session, the harness reports one context boundary, the turn end (\secref{sec:ana-temporal}).
Until it arrives, every request extends the turn's \ladenseg, which \lachesis places in HBM.
At the boundary, the harness discards that segment and materializes the \strippedseg that carries into the next turn, which \lachesis places in HBF.
In a multi-agent session, the harness also reports each sub-agent spawn, and the return that closes the sub-agent's segment (\secref{sec:ana-structural}).
The spawn separates the \subseg from the \leadseg, so from that point \lachesis places what the sub-agent writes in HBM and what the lead agent writes in HBF.
Of the two sub-agent roles, only the reviewer, whose segments are the shorter-lived, takes that placement unconditionally; a worker is admitted by the predictive path.

The predictive path ranks the workers of every session that the node serves by a lifetime predicted at their spawn, and admits the first $k$ to HBM, allocating the rest in HBF.
In a multi-agent session, every worker's segment carries the same class, because no context boundary separates one from another (\secref{sec:ana-interworker}).
They all close at sub-agent return and differ only in when that return comes, so the deterministic path cannot distinguish them.
At each worker spawn, \lachesis scores that worker's segment from its spawn signal, and the segment's place in the ranking decides its memory tier.
A reviewer is spawned only after the worker that it reviews, so \lachesis reserves HBM capacity for reviewers before it counts $k$.
$k$ is determined by HBM's free capacity rather than set as a parameter, so a fuller HBM admits fewer workers.
Only a worker's rank among the predictions is consumed, never a lifetime, so the predictor must recover the order rather than estimate how long a worker's segment lives.
Scoring the signal is one evaluation of a linear model, so it adds no delay before the worker starts.

Every memory tier decision is made in \lachesis, which holds one segment table row for each live segment.
The row is keyed by the segment id and gives the segment class, the context boundary that closes it, and the engine's own identifiers for the KV blocks that it holds.
\lachesis adds a row at a segment's first write and records in it the id of every block it allocates for that segment.
When the harness reports the boundary that a row names, \lachesis frees that row's blocks and removes the row.
The table's size is proportional to the number of blocks live at any moment.
\lachesis names the tier for every block it places, so it tracks how much KV each tier holds.

The engine executes those decisions without representing a lifetime.
\lachesis reaches the engine as two calls that it issues itself, and neither names a turn, a sub-agent, or reasoning content.
\texttt{alloc(tier)} draws one KV block from the named tier's free list, and \texttt{free([block ids])} returns the blocks that a closed segment held, each to the free list it came from.
That second call fires at a context boundary, not after the inactivity that a recency policy waits for \cite{vllm2026v1}.
A deployed engine keeps a single free block list.
Supporting a heterogeneous memory costs it one free list per tier, an HBM list and an HBF list, and nothing else.
A block's tier is therefore the free list it was drawn from.
When HBM's free list is empty, the allocation is served from the HBF free list.

%% file: contents/evaluation.tex
\section{Evaluation}
\label{sec:evaluation}

\begin{figure*}
  \centering
  \includegraphics[width=\textwidth]{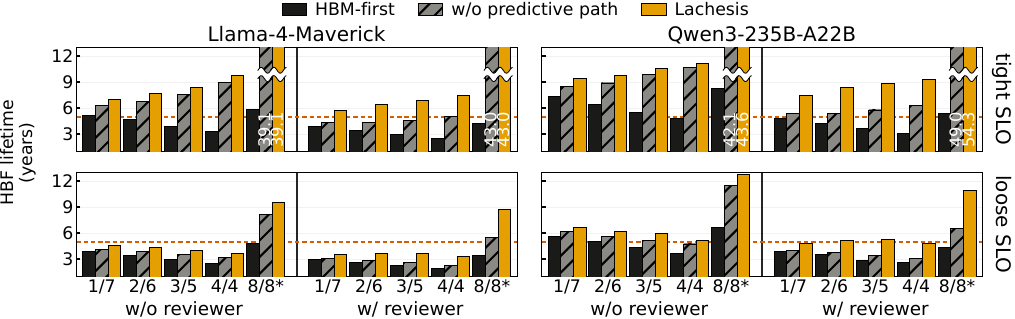}
  \caption{\placeholdervalues
  HBF lifetime in device-years on the multi-agent trace.
  The dashed line is the \rDeployPeriod warranty.}
  \label{fig:eval}
  \Description{A grid of four bar-chart panels, two rows by two columns. The columns are the two models and the rows are the two SLOs. Each panel's horizontal axis holds the multi-agent trace collected with the reviewer disabled, five GPU-memory configurations labelled w/o reviewer, then a dividing rule, then the same five with the reviewer, labelled w/ reviewer. Each block is labelled 1/7, 2/6, 3/5, 4/4 and 8/8 followed by an asterisk, evenly spaced. A bar taller than the vertical axis is drawn to the axis with a wave-shaped break across it and the value it reaches printed inside it. Every configuration in both blocks carries three bars, HBM-first, Lachesis without its predictive path, and Lachesis. The vertical axis is HBF lifetime in years, running from one to thirteen, and a dashed line marks the deployment warranty.}
\end{figure*}
\begin{figure*}
  \centering
  \includegraphics[width=\textwidth]{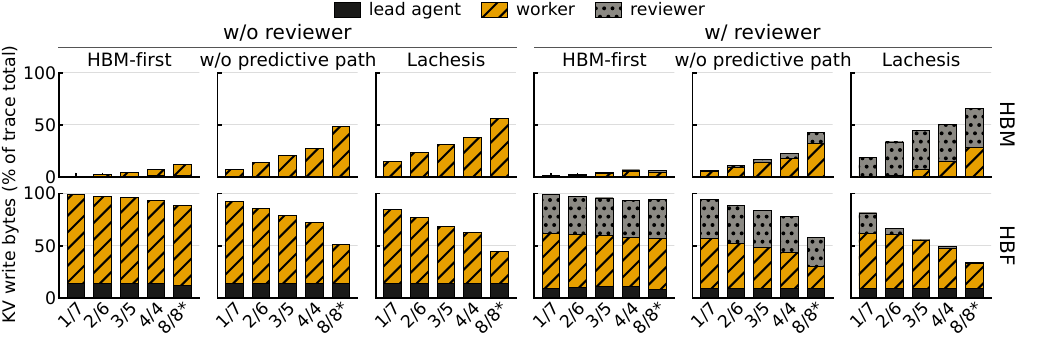}
  \caption{\placeholdervalues
    KV write bytes by segment in each tier on the multi-agent trace, with Llama-4-Maverick at the loose SLO.}
  \label{fig:writemix}
  \Description{A grid of twelve stacked-bar panels, two rows by six columns. Rows are the memory tier, HBM above HBF. The six columns fall into two groups of three, each group under a label and a horizontal rule: the run with the reviewer disabled on the left and the run as recorded on the right. Within each group the three columns are the placement policy, HBM-first, Lachesis without its predictive path, and Lachesis, so the three stand side by side at every position. Each panel carries the five GPU-memory configurations 1/7, 2/6, 3/5, 4/4 and 8/8 followed by an asterisk, their labels set diagonally, and one stacked bar at each against a shared vertical axis running from zero to one hundred percent of that run's total KV write bytes. A bar is divided into lead-agent, worker and reviewer segments under one legend, and the reviewer band is absent from the left group by construction. The two tiers at one position sum to that segment's share of the whole.}
\end{figure*}

\subsection{Implementation}
We implement \lachesis on a simulated serving engine and two-tier memory, and replay real agent traces through it.
Frontier supplies the engine \cite{feng2026frontier}: a scheduler and request queue, a KV-cache manager with a block-hash prefix cache, memory capacity expressed as a KV-block budget, and context that grows per round.
LLMSimulator supplies the hardware \cite{llmsimulator2025}, with the HBF extension prior work used \cite{son2026hbf,kyung2026hbfkv}: compiled operator kernels and roofline memory pricing.
Its memory system holds KV in both tiers and prices each read against the bandwidth of the tier it is served from.
KV bound for HBF is buffered across the decoder layers and flushed once at the end of the decode step, so HBF's write bandwidth prices that flush rather than any decoder layer.
The two meet at the operator-cost call, where the engine passes an operator's shape and its KV length and receives its execution time on the device.
Its workload generator now replays an agent run as per-request rows that name the segment each request extends.
Its KV allocator now keeps a free list per tier and serves the two calls of \secref{sec:placement}.

\subsection{Experimental Setup}
\label{sec:eval-setup}

\runin{Testbed} \lachesis is evaluated at the decode node of a prefill--decode disaggregated deployment \cite{zhong2024distserve,patel2024splitwise,feng2025windserve}, where the HBF-equipped GPUs reside \cite{son2026hbf,kyung2026hbfkv}.
Four GPU-memory configurations split the eight stack slots of the GPU used by Son et al.\ \cite{son2026hbf} between HBM and HBF, 1/7, 2/6, 3/5 and 4/4.
A fifth is H3's daisy chain, labelled 8/8* \cite{ha2026h3}.
The two tiers have the same read bandwidth and differ only in write bandwidth.
The traces are replayed against two models, Llama-4-Maverick and Qwen3-235B-A22B.
Each is served across eight GPUs, its attention by data parallelism and its experts by expert parallelism.
Every configuration is run at two SLOs on the multi-agent trace, both with a \sloTtft time to first token (TTFT) \cite{stojkovic2025dynamollm,xu2026slinfer}.
They differ in the 99th-percentile time per output token (TPOT), a tight \sloTightTpot and a loose \sloLooseTpot \cite{chen2026muxwise,stojkovic2025dynamollm}.
The single-agent traces are run at the tight SLO alone.

\runin{Workloads} Three replay traces are built from the sessions of \secref{sec:methodology} (Table~\ref{tab:regimes}).
The temporal regime supplies the two single-agent traces, $\tau^3$-bench and BFCLv4, and the structural regime's five benchmarks are mixed into one multi-agent trace.
The multi-agent trace is collected twice, once as the harness records it and once with the reviewer disabled, and every result below is read on the first unless it names the second (\secref{sec:eval-ablation}).
The evaluation uses the latency model of \secref{sec:methodology}, and both of its terms are taken from prior serving work.
No benchmark records when a user replies, so user delay is sampled from a log-normal with median \refLamUser \cite{yu2025pensieve,xie2026strata,gao2025hcache}, and tool-call latency from a published measurement for each tool class (Appendix~\ref{sec:appendix-accounting}).

\runin{Baseline} \lachesis is compared against HBM-first placement, which places a KV block in HBM whenever HBM has a free block and in HBF otherwise.
It is the placement an engine performs without \lachesis.

\runin{Metric} HBF lifetime is the time HBF's P/E budget lasts at the mean rate at which KV is written to it, excluding the one write that loads the model weights.
Each cell tolerates \hbfPE P/E cycles at SLC density \cite{son2026hbf}, over \hbfStackCap of flash per stack \cite{sandisk2025hbf}.
KV blocks are written whole and in order, so garbage collection and wear leveling move few blocks and write amplification is \hbfWaf, the value prior work adopts \cite{kyung2026hbfkv,jeong2025hifc}.
Lifetime is measured once the write traffic has settled.
A device must outlast its \rDeployPeriod warranty.

\subsection{Multi-Agent Results}
\label{sec:eval-multi}

\runin{Overall results} At a tight SLO, HBF outlasts the \rDeployPeriod warranty under \lachesis at every GPU-memory configuration and on both models, and under HBM-first placement only with Qwen3-235B-A22B on H3's daisy chain (Figure~\ref{fig:eval}, top row).
At the four configurations that give HBM one to four of the eight slots, \lachesis reaches \rEvalAbsMulti against HBM-first's \rEvalAbsMultiBase, extending HBF lifetime by \rEvalSweepGain at the tight SLO and \rEvalLifetimeLoose at the loose one.
\lachesis extends HBF lifetime over HBM-first at every model, SLO and configuration of the multi-agent trace.

\runin{What sets the gain} \lachesis's gain over HBM-first is set by one ratio, HBM's capacity against the short-lived KV live at one time, and the GPU-memory configuration and the SLO each move it.
Moving from one HBM slot in eight to four quadruples HBM capacity and removes three of the seven HBF stacks, so the aggregate P/E budget falls with them.
At the tight SLO, the added HBM absorbs \rEvalSweepBase of HBM-first's flash writes and \rEvalSweepLachesis of \lachesis's.
Only the second offsets the smaller P/E budget, so HBF lifetime falls under HBM-first and rises under \lachesis with each slot added (Figure~\ref{fig:eval}, top row).
The longest-lived configuration is therefore the smallest HBM that absorbs most of the live short-lived KV, because further slots cost more P/E budget than they save in flash writes.
A looser SLO instead admits more concurrent sessions, so that ratio falls and the gain narrows (Figure~\ref{fig:eval}, bottom row).
It also moves the longest-lived configuration down to two or three HBM slots in eight.

\runin{Write composition} \lachesis writes \rEvalWriteMixGain as much KV to HBM as HBM-first placement, so HBF receives less and spends less of its P/E budget (Figure~\ref{fig:writemix}).
HBM takes \rEvalWriteMixHbmLachesis of a run's write bytes under \lachesis against \rEvalWriteMixHbmBase under HBM-first, and the rest of that volume reaches HBF.
HBM-first frees blocks only under memory pressure, so its HBM capacity does not turn over.
\lachesis places the \leadseg in HBF at the write and admits the workers predicted to be short-lived, so HBM's capacity turns over.
The reviewer increases the short-lived KV live at one time, and HBM's capacity does not grow with it, so more of that KV reaches HBF and HBF lifetime falls under both placement policies.
\lachesis places that KV in HBM by its segment class, so the loss is smaller under \lachesis than under HBM-first (\secref{sec:eval-ablation}).

\runin{H3's daisy chain} \lachesis extends HBF lifetime by \rEvalHthreeTight at the tight SLO on H3's daisy chain, the largest gain at any configuration.
H3 does not trade HBM capacity against flash stacks, because its eight slots carry HBM and HBF together instead of dividing them, so HBM grows and no flash stack is lost (Figure~\ref{fig:eval}, 8/8*).
The gain falls to \rEvalHthreeLoose on the change of SLO alone.

\subsection{Ablation Study}
\label{sec:eval-ablation}

\lachesis's gain over HBM-first is the product of two margins, one from each placement path (\secref{sec:eval-multi}).
\lachesis \emph{without the predictive path} places every segment by the segment class the harness fixes and ranks no worker.
It yields \rEvalAblationRule over HBM-first on the multi-agent trace, and \lachesis adds \rEvalAblationRank over it (Figure~\ref{fig:eval}).
Both margins hold on both models and at every configuration that splits the eight slots.

\runin{What the reviewer changes} The predictive path ranks only worker segments, and what it earns is set by the HBM capacity that ranking keeps free for the reviewer, \rEvalAblationRank with a reviewer present against \rEvalNoReviewer without one (Figure~\ref{fig:eval}).
A reviewer's segment is generally shorter-lived than a worker's, so \lachesis places it in HBM by its segment class with or without the predictive path (\secref{sec:placement}).
Without the ranking, long-lived worker segments enter HBM and hold its capacity until their workers return.
\lachesis instead admits the worker segments predicted to be short-lived, which are also the ones that write less (\secref{sec:ana-interworker}), so the same HBM covers more of the short-lived KV live at one time and usually has room for a reviewer.
The reviewer's own write volume reaches flash at \rEvalReviewerSpillRule without the predictive path, and at \rEvalReviewerSpillLachesis under \lachesis (Figure~\ref{fig:writemix}).
In the sessions with no reviewer, workers are the only segments HBM holds, so ranking them only changes which of them it holds.

\subsection{Single-Agent Results}
\label{sec:eval-single}

\runin{Overall results} HBF outlasts the \rDeployPeriod warranty under \lachesis at every configuration but one, and under HBM-first placement at some of $\tau^3$-bench's and none of BFCLv4's (Figure~\ref{fig:singleagent}).
At the four configurations that give HBM one to four of the eight slots, \lachesis reaches \rEvalAbsSingle against HBM-first's \rEvalAbsSingleBase, extending HBF lifetime by \rEvalLifetimeSingle.
These traces are single-agent, so they exercise the temporal axis alone and the deterministic path places every segment (\secref{sec:placement}).

\runin{What a larger HBM admits} \lachesis's lifetime peaks at one to three HBM slots in eight and falls with every slot added past the peak (Figure~\ref{fig:singleagent}).
A larger HBM admits no more KV once it holds every \ladenseg the deployment has live at one time, and its share of a run's write bytes flattens (Figure~\ref{fig:singlewritemix}).
Past the peak, every added slot costs an HBF stack and saves no flash write, and \lachesis reaches \rEvalSingleAbsFall fewer device-years on BFCLv4 at four HBM slots in eight than at one.
Under \lachesis, H3's daisy chain loses no flash stack and reaches the longest lifetime at every benchmark and model.
The multi-agent trace peaks later at the same SLO, because its worker segments give \lachesis more to admit than HBM has room for (\secref{sec:eval-multi}).

%% file: contents/related.tex
\section{Related Work}
\label{sec:related}

\lachesis places KV in flash by the lifetime of a context segment, which no HBF design for LLM inference has used to decide what flash holds.
Early designs place only read-only data.
H3 \cite{ha2026h3} stores model weights and precomputed KV in HBF and keeps generated KV in HBM, and FLINT \cite{oliveira2026flint} replaces the flash translation layer with a read-only one, which locates model weights and supports no writes.
Neither design spends flash endurance on generated KV, because neither writes it.
\begin{figure}
  \centering
  \includegraphics[width=\columnwidth]{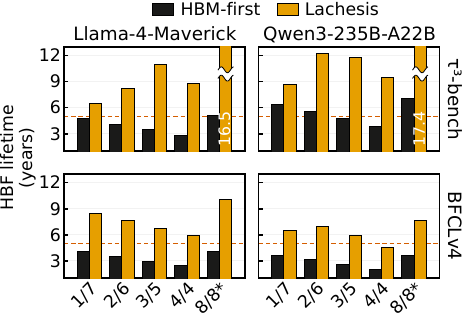}
  \caption{\placeholdervalues
  HBF lifetime in device-years on the single-agent traces.
  The dashed line is the \rDeployPeriod warranty.}
  \label{fig:singleagent}
  \Description{A grid of four bar-chart panels, two rows by two columns. Rows are the two single-agent benchmarks, $\tau^3$-bench above BFCLv4, and columns are the two models. Each panel's horizontal axis holds the five GPU-memory configurations, with two bars at each --- HBM-first and Lachesis --- against a vertical axis of HBF lifetime in years, running from one to thirteen, and a dashed line marking the deployment warranty. A bar taller than the axis is drawn to it with a wave-shaped break and the value it reaches printed inside. The configurations are labelled 1/7, 2/6, 3/5, 4/4 and 8/8 followed by an asterisk, evenly spaced.}
\end{figure}
\begin{figure}
  \centering
  \includegraphics[width=\columnwidth]{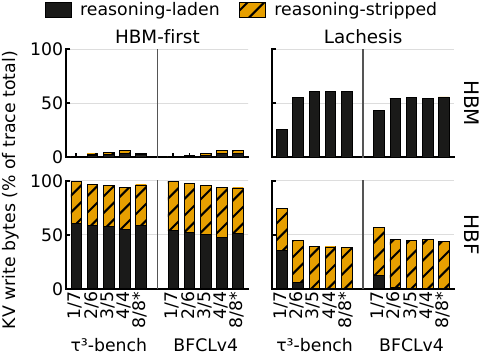}
  \caption{\placeholdervalues
  KV write bytes by segment in each tier on the single-agent traces, with Qwen3-235B-A22B at the tight SLO.}
  \label{fig:singlewritemix}
  \Description{A grid of four stacked-bar panels, two rows by two columns. Rows are the memory tier, HBM above HBF, and columns are the placement policy, HBM-first and Lachesis. Each panel is divided by a vertical rule into two blocks, $\tau^3$-bench on the left and BFCLv4 on the right, and each block carries the five GPU-memory configurations 1/7, 2/6, 3/5, 4/4 and 8/8 followed by an asterisk, their labels set vertically. Every configuration carries one stacked bar against a shared vertical axis running from zero to one hundred percent of that run's total KV write bytes, divided into reasoning-laden and reasoning-stripped under one legend.}
\end{figure}
Later designs do write generated KV, and each answers the write cost with a mechanism.
FlashAccel \cite{wang2026flashaccel} places weights and KV together in flash, Park et al.\ \cite{park2026hbm} offload a buffered write-back into a custom base die, and DASH \cite{kim2026beyondcapacity} holds decode KV in HBM until it can be written to flash in whole pages.
These designs lower what each write costs, and DASH also decides where data goes, from its size, mutability and reuse.
None of the three decides from how long the data stays readable.
Son et al.\ \cite{son2026hbf} measure what writing every generated KV block to flash costs.
When HBF serves as the main GPU memory, each flash block accumulates more P/E cycles than SLC NAND tolerates.
Kyung et al.\ \cite{kyung2026hbfkv} show that one property can decide which KV is worth that cost, a request's output length (\secref{sec:bg-why-lifetime}).
\lachesis instead decides from a lifetime the agent harness fixes before the segment is written.

%% file: contents/conclusion.tex
\section{Conclusion}
\label{sec:conclusion}

This paper presents \lachesis, a lifetime-aware KV cache placement layer for agent serving across HBM and HBF.
We find that the context an agent session accumulates divides at harness-defined context boundaries into context segments whose lifetimes differ along three axes, temporal, structural and \interworker, and that each lifetime is knowable before its segment is written.
\lachesis places each segment on the tier its lifetime implies, along a deterministic path that reads the segment class the harness fixes and a predictive path that ranks worker segments at spawn.
In trace-driven simulation, \lachesis extends HBF lifetime by \rEvalHbfLifetime over HBM-first placement, and HBF outlasts a \rDeployPeriod warranty on the multi-agent trace at a tight SLO even under 24$\times$7 operation.

\section*{Acknowledgments}
This paper was the result of the research project supported by SK hynix Inc. Jae W. Lee is the corresponding author.
This work used Claude Code for code, language editing and text refinement. All technical content, experiments, and conclusions were verified by the authors.

%% file: contents/appendix_corpus.tex
\section{Workload Corpus Statistics}
\label{sec:appendix-corpus}

\runin{Harness configuration} The temporal regime runs each benchmark under Harbor's own official prompts.
Each structural-regime benchmark's official prompts and reference agent implementation are ported to AOrchestra.

\runin{Token columns} Tokens processed counts every request's input and output, so the context a round reuses is counted again at each round.
Tokens written counts only what each request newly materializes, which is the quantity \secref{sec:ana-temporal} reports write volume over.
The structural traces record per-segment sizes rather than per-request lengths, so the table reports only the second column for them.

\input{tables/t4_corpus}

No session is excluded from either corpus, including those that failed or timed out.

%% file: tables/t4_corpus.tex
\begin{table*}[t]
  \centering
  \small
  \caption{Corpus scale per workload.}
  \label{tab:corpus}
  \begin{tabularx}{\textwidth}{L{0.7}L{1.3}rrrr}
    \toprule
    Regime & Workload & Sessions & Requests & Tokens processed & Tokens written \\
    \midrule
    Temporal
      & $\tau^3$-bench airline      &  58 &    709 &  4{,}735{,}468 &   591{,}248 \\
      & $\tau^3$-bench retail       & 114 &  1{,}387 &  9{,}593{,}474 & 1{,}156{,}112 \\
      & $\tau^3$-bench telecom      & 114 &  2{,}208 & 20{,}687{,}494 & 1{,}407{,}427 \\
      & $\tau^3$-bench banking      &  97 &  2{,}221 & 63{,}759{,}218 & 7{,}172{,}200 \\
      & BFCLv4 base                 & 200 &  2{,}063 &  8{,}865{,}141 &   800{,}193 \\
      & BFCLv4 long-context         & 200 &  2{,}066 & 23{,}312{,}059 & 5{,}279{,}265 \\
    \addlinespace
    Structural
      & AppWorld                    & 168 &  6{,}248 & --- &  9{,}738{,}267 \\
      & DeepSearchQA                & 125 & 12{,}241 & --- & 37{,}196{,}143 \\
      & SWE-bench Verified          & 149 &  8{,}745 & --- & 11{,}677{,}326 \\
      & TerminalBench               &  75 &  3{,}675 & --- &  7{,}289{,}834 \\
      & WebArena                    & 106 &  2{,}190 & --- &  6{,}707{,}151 \\
    \bottomrule
  \end{tabularx}
\end{table*}

%% file: contents/appendix_accounting.tex
\section{Trace Fields, Clock Charges, and KV Accounting}
\label{sec:appendix-accounting}

This appendix gives the trace fields, the clock charges and the exact $\rhorwbare$ accounting that \secref{sec:methodology} states in one sentence each, for reproducibility \cite{kyung2026hbfkv}.

The trace records, for each request, its session, turn, and sub-agent lineage, its tokenized input and output, and its tool-call and observation boundaries, alongside the two cache fields \secref{sec:methodology} names.
With the trace aligned in session order, the matched prefix gives the KV reused from earlier requests, and the unmatched suffix plus the generated tokens gives the KV newly materialized.
A segment built over several requests becomes live at its first write.

Appending a tool result forces a partial prefill before the next round can decode.
The clock of \secref{sec:methodology} charges $\ttftref$ once per observation, whatever that observation's length.
A prefill deployment provisioned to the reference point completes the prefill for any single observation within that time.
$\lamtool$ is drawn from an exponential distribution \cite{ye2026speculative} whose mean is the published measurement for the calling tool's class \cite{abhyankar2024infercept}, at the granularity each regime's trace records tool identity.

Let $\Vkvtok$ be the per-token KV footprint of the evaluated attention architecture, and let $\Nw$ be the number of tokens whose KV state is materialized for segment $\seg$ at the decode node.
Then
\begin{equation}
  \Vw = \Nw \, \Vkvtok .
  \label{eq:write-volume}
\end{equation}

For request $q$ with output length $\Lout(q)$, let $n_{\seg}(q,i)$ be the number of KV tokens from $\seg$ present before its $i$-th decode step.
Under full-context attention,
\begin{equation}
  \Vr = \Vkvtok \sum_{q \in \Qset} \sum_{i=1}^{\Lout(q)} n_{\seg}(q,i),
  \qquad
  \rhorw = \frac{\Vr}{\Vw},
  \label{eq:read-volume}
\end{equation}
where $\Qset$ is the set of requests whose decode stage reads $\seg$.

For a stateless request this reduces to Kyung et al.'s request-level expression, under their decode-stage accounting but computed per segment.
Only decode-stage reads count: observation-induced partial prefill is counted in $\tnormbare$ through the $\ttftref$ term, not $\Vr$.
Initial prefill tokens are counted in $\Vw$, while initial prefill time is not counted in $\tnormbare$.